\documentstyle[psfig,aps,prl,multicol]{revtex}
\begin{document}
\title{Optical Simulation of Quantum Logic}
\author{N. J. Cerf$^1$, C. Adami$^{1,2}$, and P. G. Kwiat$^3$}
\address{$^1$W. K. Kellogg Radiation Laboratory and 
         $^2$Computation and Neural Systems\\
California Institute of Technology,
Pasadena, California 91125\\
         $^3$Physics Division, P-23, Los Alamos National Laboratory,
Los Alamos, New Mexico 87545}

\date{March 1997}

\draft
\maketitle
\vskip -0.1cm

\begin{abstract}
A systematic method for simulating small-scale quantum circuits by use of
linear optical devices is presented. It relies on the representation
of several quantum bits by a single photon, and on the implementation
of universal quantum gates using simple optical components
(beam splitters, phase shifters, etc.). This suggests that 
the optical realization of small quantum networks
is reasonable given the present technology in quantum optics, and could
be a useful technique for testing simple quantum algorithms
or error-correction schemes.
The optical circuit for quantum teleportation is presented
as an illustration.
\end{abstract}

\pacs{PACS numbers: 03.65.Bz,42.50.-p,42.79.Ta,89.70.+c
      \hfill KRL preprint MAP-211}

\begin{multicols}{2}[]
\narrowtext

Quantum computation can be described as the task of performing 
a specific unitary transformation on a set of quantum bits (qubits) 
followed by measurement, so that the outcome of the
measurement provides the result of the computation.
This unitary transformation can be constructed with a finite number
of $2\times 2$ unitary matrices, that is, using a quantum circuit utilizing
only 1-bit and 2-bit quantum gates (see, {\it e.g.},
\cite{bib_qcircuit,bib_barenco}). The universality of 1- and 2-bit
gates in the realization of an arbitrary quantum computation has been proven
in~\cite{bib_barenco,bib_universal}.
It has been shown recently that an optical realization
exists for any $N\times N$ unitary matrix~\cite{bib_reck}, a result which
generalizes the well-known implementation of $U(2)$ matrices
using a lossless beam splitter and a phase shifter
(see, {\it e.g.}, \cite{bib_U2}). Accordingly,
each element of $U(N)$ can be constructed using
an array of ${\cal O}(N^2)$ beam splitters that form an optical multiport
with $N$ input and $N$ output beams.
In this note, we focus on the simulation of universal
quantum gates using {\em linear} optics components,
and propose a systematic method to assemble 
these optically-simulated gates to build simple quantum circuits.
\par

\begin{figure}
\caption{Example of optical simulation of basic quantum logic
gates. (a) Hadamard gate on a ``location'' qubit, using a lossless
symmetric beam splitter.
(b) Controlled-{\sc not} gate using a polarization rotator.
The location and polarization
are the control and target qubit, respectively. (c) Same as (b) but
the control and target qubits are interchanged by the use of a polarizing
beam splitter.}
\vskip 0.25cm
\centerline{\psfig{figure=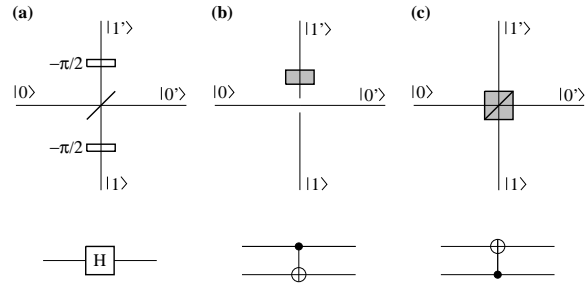,width=3.0in,angle=-90}}
\label{fig_intro}
\vskip -0.25cm
\end{figure}

In what follows, we discuss a {\em correspondence}
between quantum networks and 
linear optical setups, and present as an example the optical
realization of a 3-bit quantum computation. This is achieved by introducing
a {\em single-photon} representation
of several quantum bits, building on the equivalence between
traditional linear optics elements (such as beam splitters or
phase shifters) and 1-bit quantum gates (see, {\it e.g.}, \cite{bib_chuang}).
For example, in quantum circuit terminology,
an optical symmetric beam splitter is known to act
as a quantum $\sqrt{\rm \sc not}$ gate (up to a phase of $\pi/4$)
if we use the pair of input modes $|01\rangle$ (or $|10\rangle$) to represent
the logical 0 (or 1) state of the qubit. If one input port is in 
the vacuum state $|0\rangle$ and the second one in a single-photon
state $|1\rangle$, the output ports will then be in a superposition state 
$|01\rangle +i |10\rangle$. Similarly, a quantum phase gate
can be obtained by use of a phase shifter acting on one mode of the photon.
In other words, single-photon interferometry experiments 
can be interpreted in quantum circuit language,
the ``which-path'' variable being substituted with a quantum bit.
Although a general constructive proof for the existence of an optical
realization of an arbitrary quantum circuit is implicitly given 
in Ref.~\cite{bib_reck}, the simple duality between 
quantum logic and single-photon optical experiments is not exploited.
Here, we use the fact that several (say $n$) quantum bits can be represented 
by a {\em single} photon in an interferometric setup involving
essentially $2^n$ paths, so that conditional dynamics can
easily be implemented by using different optical elements in distinct paths.
The appropriate cascading of beam splitters and other linear optical
devices entails the possibility of simulating 
networks of 1- and 2-bit quantum gates
(such as the Hadamard or the controlled-{\sc not} gate, 
see Fig.~\ref{fig_intro}), and thereby in principle achieving
universal $n$-bit quantum computations.
This is in contrast with traditional optical models
of quantum logic, where in general $n$ photons interacting
through nonlinear devices (acting as 2-bit quantum gates)
are required to represent $n$ qubits (see, {\it e.g.}, \cite{bib_chuang}).
Such models typically make use of the Kerr nonlinearity
to produce intensity-dependent phase shifts, so that the presence of
a photon in one path induces a phase shift to a second photon
(see, {\it e.g.}, the optical realization of a Fredkin 
gate~\cite{bib_milburn}). Instead, the model proposed here
yields a straightforward method for ``translating'' an $n$-bit
quantum circuit into a single-photon optical setup, 
whenever $n$ is not too large. The price to pay
is the exponential growth of the number of optical paths,
and, consequently, of optical devices that are required.
This will most likely limit the applicability of the proposed technique
to the simulation of relatively simple circuits. 
As an illustration, we show that the quantum
circuit for teleportation (involving 3 qubits and 8 quantum gates,
see~\cite{bib_teleport}) can be simulated optically 
using essentially 9 beam splitters.
\par

First, let us consider a single-photon experiment
with a Mach-Zehnder interferometer 
in order to illustrate the optical simulation of elementary quantum gates
(see Fig.~\ref{fig_intro}).
One qubit is involved in the description of the interferometer
in terms of a quantum circuit: the ``location'' qubit, characterizing
the information about ``which path'' is taken by the
photon. Rather than using the occupation number representation
for the photon, here we label the two input modes entering the
beam splitter by $|0\rangle$ and $|1\rangle$ (``mode description''
representation). The quantum state of the photon {\em exiting}
the beam splitter then is $|0'\rangle + i |1'\rangle$ or  
$|1'\rangle + i |0'\rangle$ depending on the input mode of the photon.
(The factor $i$ arises from the $\pi/2$ phase shift between
the transmitted and the reflected wave in a 
lossless symmetric beam splitter~\cite{bib_pi/2}.) This is
the $\sqrt{\rm \sc not}$ gate discussed earlier. Placing phase 
shifters at the input and output ports as shown in Fig.~\ref{fig_intro}a,
the beam splitter can be shown to perform
a Hadamard transformation between input and output modes, {\it i.e.},
\begin{equation}
\left( \begin{array}{c} |0'\rangle \\ |1'\rangle \end{array} \right)
= {1 \over \sqrt{2}}
\left( \begin{array} {cr} 1 & 1 \\ 1 & -1 \end{array} \right)
\left( \begin{array}{c} |0\rangle \\ |1\rangle \end{array} \right)  \;.
\end{equation}
In this sense, a lossless symmetric beam splitter (supplemented
with two $-\pi/2$ phase shifters) can be viewed as a Hadamard gate
acting on a location qubit.
Recombining the two beams using a second beam splitter
in order to form a balanced Mach-Zehnder interferometer corresponds 
therefore, in this quantum circuit language, 
to having a second Hadamard gate acting subsequently 
on the qubit~\cite{fn_dynphases}.
Since $H^2=1$, it is not a surprise that the location qubit returns
to the initial basis state ($|0\rangle$ or $|1\rangle$) 
after two beam splitters. This simple quantum circuit
(a sequence of two Hadamard gates) 
therefore describes the fact that the contributions
of the two paths interfere destructively in one of the output ports, so that
the photon always leaves the interferometer in the same direction
as it entered.
\par 

More interestingly,
consider now the same interferometer using polarized photons
(the photon is horizontally polarized at the input). Assuming that none
of the devices acts on polarization, the photon exits the interferometer
with the same polarization. In a circuit language, 
this corresponds to introducing
a ``polarization'' qubit ($|0\rangle$ stands for horizontal polarization)
which remains in a product state with the location qubit
throughout the circuit. If a polarization
rotator is placed in one of the branches of the interferometer, flipping
the polarization from horizontal $|0\rangle$ to vertical $|1\rangle$,
it is well known that interference disappears since both paths
become distinguishable. This corresponds to placing
a 2-bit controlled-{\sc not} gate (represented in Fig.~\ref{fig_intro}b)
between the two Hadamard gates, where
the location qubit is the control and polarization is the target bit.
Conditional dynamics is achieved in the sense that the polarization
of the photon is flipped conditionally on its location. The
disappearance of interference then simply reflects the entanglement between
location and polarization qubits (the reduced density matrix 
obtained by tracing over polarization shows that the photon ends up in
is a mixed ``location'' state, {\it i.e.}, it
has a 50\% chance of being detected in one or the other exit port).
According to this, Feynman's rule of thumb (namely that interference and
which-path information are complementary) is a manifestation
of the quantum {\em no-cloning} theorem: the location qubit cannot be
``cloned'' into a polarization qubit.
\par

The optical equivalent of other basic quantum gates can be devised
following the same lines. For example, a polarizing beam splitter
achieves a controlled-{\sc not} gate where the location qubit is
flipped or not (the photon is reflected or not) conditionally on its
state of polarization, as shown in Fig.~\ref{fig_intro}c. 
Fredkin, Toffoli, as well as
controlled-phase gates can easily be simulated in the same manner but
will not be considered here. The central point is that, in principle, a
universal quantum computation can be simulated using these optical
substitutes for the universal quantum gates. The optical setup
is constructed straightforwardly by inspection of the quantum circuit.
A circuit involving $n$ qubits requires in
general $n$ successive splitting stages of the incoming beam,
that is, $2^n$ optical paths are obtained via $2^n -1$ beam splitters.
(The use of polarization of the photon as a qubit allows using
$2^{n-1}$ paths only.) This technique is thus limited to the simulation
of quantum networks involving a relatively small number of qubits
(say less than 5-6 with present technology).
The key idea of a quantum
computer, however, is to avoid just such an exponential size of the
apparatus by having $n$ physical qubits performing unitary transformations
in a $2^n$-dimensional space. In this respect,
it can be argued 
that an optical setup requiring $\sim 2^n$ optical 
elements to perform an $n$-bit quantum computation represents
a {\em classical} optical computer (see, {\it e.g.}, \cite{bib_barenco}).
Rather than debating this issue, our intention here
is to show how to {\em simulate} small-$n$ quantum circuits
using standard linear optics, which should prove to be
useful for testing experimentally non-trivial quantum
circuits or simple quantum algorithms.
\par

\begin{figure}
\caption{Quantum circuit for teleportation 
(from~\protect\cite{bib_teleport}). The initial state of qubit $\Lambda$ is
teleported to the state of qubit $\lambda$. Qubits $\sigma$ and $\lambda$
must be initially in state $|0\rangle$. Qubits $\Lambda$ and $\sigma$,
if measured at the end of the circuit, yield two classical (random)
bits that are uniformly distributed. }
\vskip 0.25cm
\centerline{\psfig{figure=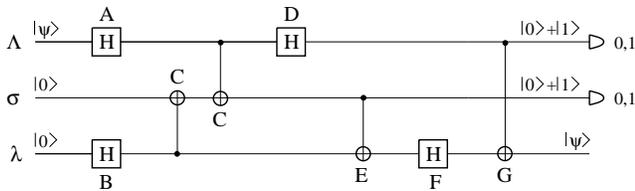,width=3.3in,angle=-90}}
\label{fig_circuit}
\vskip -0.25cm
\end{figure}

Let us focus on the quantum circuit for teleportation~\cite{bib_teleport} 
shown in Fig.~\ref{fig_circuit}. 
This circuit has the property that the arbitrary
initial state $| \psi \rangle$ of qubit $\Lambda$ 
is teleported to the state in which qubit $\lambda$
is left after the process. In the original teleportation 
scheme~\cite{bib_originalteleport}, two classical bits (resulting from a Bell
measurement) are sent by the emitter, while the receiver performs
a specific unitary operation on $\lambda$ depending on these two bits.
However, it is shown in \cite{bib_teleport} that these unitary operations
can be performed at the quantum level as well, by using quantum logic gates 
and postponing the measurement of the two bits to the end of the circuit.
The resulting quantum circuit (Fig.~\ref{fig_circuit})
retains the essence of teleportation. 
\par

\begin{figure}
\caption{Optical realization of the quantum circuit for teleportation
using polarized photons. The location qubit $\Lambda$ characterizes
the ``which-arm'' information at the first beam splitter, while
qubit $\lambda$ stands for the ``which-path'' information at the
second level of splitting. The initial location qubit $\Lambda$ is teleported
to qubit $\lambda$, probed via the interference pattern observed at the upper
or lower ($\Lambda=0,1$) final beam splitter, for both polarization 
states ($\sigma=0,1$) of the detected photon. }
\vskip 0.25cm
\centerline{\psfig{figure=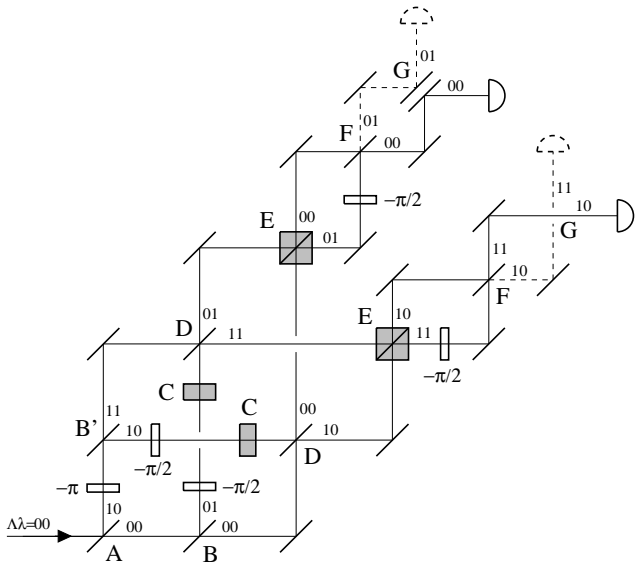,width=3.3in,angle=-90}}
\label{fig_optics}
\vskip -0.25cm
\end{figure}

In our proposed optical realization
of this circuit (see Fig.~\ref{fig_optics}),
qubits $\Lambda$ and $\lambda$ correspond to the location of the photon
at the first and second splitting level, while $\sigma$
stands for the polarization qubit. Note that
the photons are initially horizontally
polarized, {\it i.e.}, in polarization state $|0\rangle$.
The first beam splitter A in Fig.~\ref{fig_optics} acts as a Hadamard gate
on $\Lambda$, as explained previously. 
For convenience, we depict the teleportation of state
$|\psi\rangle=|0\rangle$, so that
the incident photon enters this beam splitter in the input port
labeled $|0\rangle$. However, as any operation
in $U(2)$ can be realized optically, an {\em arbitrary} state of $\Lambda$
can be prepared (and then teleported) by having an initial
beam splitter (with appropriate phase shifters) connected to
both input ports of beam splitter A.
The second level of beam splitters B (and B'~\cite{fn_bsp}) corresponds
to the Hadamard gate B on $\lambda$ in Fig.~\ref{fig_circuit}.
The four paths at this point
($\Lambda \lambda =00$, 01, 10, and 11) label the four
components of the state vector characterizing qubits $\Lambda$ and $\lambda$. 
The probability amplitude for observing the photon in each of these
four paths, given the fact the photon enters the $|0\rangle$
port of beam splitters A and B, is then simply the corresponding
component of the wave vector. The combined
action of both controlled-{\sc not} gates C in Fig.~\ref{fig_circuit}
is to flip the
polarization state of the photon (qubit $\sigma$) conditionally on
the parity of $\Lambda+\lambda$ (modulo 2), which is achieved by inserting
polarization rotators C at the appropriate positions. In other words,
the polarization is flipped on path 01 or 10, while it is unchanged
on path 00 or 11. 
\par

The Hadamard gate D in Fig.~\ref{fig_circuit}
acts on qubit $\Lambda$, independently of $\lambda$. This is achieved
in Fig.~\ref{fig_optics}
by grouping the paths in pairs with the same value of $\lambda$
({\it i.e.}, crossing the paths) and using two beam
splitters D in order to effect a Hadamard transformation
on $\Lambda$ (one for each value of $\lambda$).
Similarly, the controlled-{\sc not} gate E acting on $\lambda$ (conditionally
on the polarization) is simulated by the use of two polarizing beam
splitters E after crossing the paths again~\cite{fn_polbeamsplitter}.
The last Hadamard gate F in Fig.~\ref{fig_circuit}
corresponds to the two last
beam splitters F, and the final controlled-{\sc not} gate G
is simply achieved by crossing the paths ($\lambda=0,1$) in the lower arm
($\Lambda=1$) versus the upper arm ($\Lambda=0$).
In fact, the setup could be simplified by noting that
the conditional crossing of paths achieved by G simply reduces to
relabeling the output ports of beam splitter F in the $\Lambda=1$ arm.
In Fig.~\ref{fig_optics}, only those phase shifters associated
with the Hadamard gates (Fig.~\ref{fig_intro}a) are indicated
that are relevant in the final detection.
\par

The interpretation of this optical circuit in terms of teleportation
is the following. After being ``processed'' in this quantum circuit,
a photon which was initially horizontally polarized can reach
one of the two ``light'' detectors (solid line in Fig.~\ref{fig_optics}) 
with horizontal or vertical polarization.
This corresponds to the final measurement of qubits $\Lambda$ and
$\sigma$ in Fig.~\ref{fig_circuit},
yielding two classical (random) bits: upper or lower arm,
horizontal or vertical polarization. The third qubit, $\lambda$, contains
the teleported quantum bit, that is, the initial arbitrary state of $\Lambda$.
Since the location state of the photon is initially $|0\rangle$
in the setup represented in Fig.~\ref{fig_optics},
it always exits to the ``light'' detector and never reaches the ``dark''
one (dashed line). For any measured value of $\Lambda$
(photon detected in the upper or lower arm) and $\sigma$ (horizontally
or vertically polarized photon), the entire setup forms a simple
balanced Mach-Zehnder interferometer. Indeed, there are exactly two
{\em indistinguishable} paths leading to each of the eight possible
outcomes (four detectors, two polarizations); these interfere pairwise,
just as in a standard Mach-Zehnder interferometer, explaining
the fact that the photon always reaches the ``light'' detector
(in both $\Lambda=0$ and $\Lambda=1$ arms and for both polarizations).
In this sense, the initial
``which-arm'' qubit $\Lambda$ has been teleported to the final
``which path'' qubit $\lambda$. 
This process is {\em formally} equivalent
to the original teleportation scheme~\cite{bib_originalteleport}
(although no classical bits are communicated) as
exactly the same unitary transformations and quantum gates are involved.
Note also that, as no
photodetection coincidence is required in this optical experiment,
the setup is actually not limited to {\em single}-photon interferometry.
This largely simplifies the realization of the optical source
since classical light fields (such as those from a laser) can be used rather
than number states.
\par

An actual experimental realization of the setup in Fig.~\ref{fig_optics}
should be
straightforward, if non-trivial.  First, in order to avoid unwanted
polarization effects at any of the mirrors and non-polarizing beam
splitters, one would want to arrange the optics so that the various
reflections occurred at near-normal incidence (thereby removing the
distinction between {\it s} and {\it p} polarizations).
The main difficulty in the
setup is that various path lengths in the system should be the same.  This
could be achieved by adjusting for white-light fringes in each of the
sub-interferometers ({\it e.g.}, the Mach-Zehnder interferometer formed by the
beam splitters A and lower D; the interferometer formed by the beam splitters
B and upper E, etc.), without the additional phase shifters or polarization
rotators.  These latter elements could then be ``added'' by simply rotating
appropriate birefringent wave plates already in the system.  For example, an
exact $\pi$ phase-shift is produced by simply rotating the slow axis of a
half-wave retardation plate from horizontal to vertical~\cite{fn_pol};
a 90$^\circ$ polarization flip is caused by
rotating a half-wave plate from horizontal to 45$^\circ$. When each of the
subsystems is properly adjusted, and the extra phase and
polarization-rotation elements correctly set, the entire system should
perform as indicated, i.e., a photon incident from the left should only
exit via the right-directed output ports. Finally, a tunable input
$|\psi\rangle = a |0\rangle + b |1\rangle$ 
(obtained with an additional beam splitter
at the input) can be used to verify that any arbitrary state
is faithfully teleported.

\par

We have proposed a general technique for simulating
small-scale quantum networks using optical setups
composed of linear optical elements. This avoids the recourse to
non-linear Kerr media to effect quantum conditional dynamics,
a severe constraint in the usual optical realization of quantum circuits.
A drawback of this technique
is clearly the exponential increase of the resources (optical devices)
with the size of the circuit. 
Nevertheless, as optical components that simulate
1- and 2-bit universal quantum gates can be cascaded straightforwardly,
a non-trivial quantum computing optical device can be constructed
if the number of component qubits is not too large. 
We believe this technique can be applied without fundamental difficulties
to the encoding and decoding circuits that are involved in the
simplest quantum error-correcting schemes~\cite{bib_qec}, 
opening up the possibility for an experimental simulation of them.
\par
 
This work was supported in part by NSF Grants
PHY 94-12818 and PHY 94-20470, and by a grant from DARPA/ARO
through the QUIC Program (\#DAAH04-96-1-3086).

\end{multicols}

\begin{references}
\bibitem{fn_dynphases}Here and below, 
it is understood that the path lengths are
adjusted so that the difference between dynamical phases vanishes.
\bibitem{fn_bsp} For convenience,
two realizations (B and B') of the Hadamard gate are used 
in Fig.~\ref{fig_optics}, where B' is obtained from B by interchanging
the $|0'\rangle$ and $|1'\rangle$ output ports
in Fig.~\ref{fig_intro}a.
\bibitem{fn_polbeamsplitter} A polarizing beam splitter
leaves an horizontally polarized photon (in polarization
state $|0\rangle$) unchanged, while vertical
polarization (state $|1\rangle$) is reflected.
\bibitem{fn_pol} Phase
shifts that occur after polarization transformations C (and which must
therefore be independent of polarization) could be produced by a calibrated
rotation of a thin glass plate.
\end{references}
\end{document}